\documentclass{elsart}

\usepackage{graphicx}
\usepackage{amssymb,amsbsy}
\def\be{\begin{equation}}
\def\ee{\end{equation}}
\def\bea{\begin{eqnarray}}
\def\eea{\end{eqnarray}}
\def\bear{\begin{array}}
\def\ear{\end{array}}
\def\bfig{\begin{figure}}
\def\efig{\end{figure}}
\def\bcen{\begin{center}}
\def\ecen{\end{center}}
\def\raw{\rightarrow}

\def\Bra#1{\bigl\langle #1\bigr|}
\def\Ket#1{\bigl| #1\bigr\rangle}
\def\vp{\mathbf{p}}
\def\vk{\mathbf{k}}

\def\vq{\mathbf{q}}
\def\mpi{m_{\pi}}
\def\la{\label}
\def\chic{\scriptscriptstyle}

\def\D{\displaystyle}
 
\def\bkappa{\boldsymbol\kappa}
\def\boeta{\boldsymbol\eta}

\begin{document}

\begin{frontmatter}

\title{Nuclear response functions for the $\mathbf{N-N^*(1440)}$ transition}

\author[dire1]{L. Alvarez-Ruso}, \author[dire2]{M. B. Barbaro}, 
\author[dire3]{T. W. Donnelly} and \author[dire2]{A. Molinari}

\address[dire1]{Instit\"ut f\"ur Theoretische Physik, Universit\"at Giessen,\\
D-35392 Giessen, Germany}
\address[dire2]{Dipartimento di Fisica Teorica,
Universit\`a di Torino \\and
INFN, Sezione di Torino \\
Via P. Giuria 1, 10125 Torino, Italy}
\address[dire3]{Center for Theoretical Physics, \\
Laboratory for Nuclear Science and Department of Physics\\
Massachusetts Institute of Technology,
Cambridge, MA 02139, USA}

\begin{abstract}
Parity-conserving and -violating response 
functions are computed for the inclusive electroexcitation of the $N^*(1440)$
(Roper) resonance in nuclear matter modeled as a relativistic Fermi
gas. Using various empirical parameterizations and
theoretical models of the $N-N^*(1440)$ transition form factors,    
the sensitivity of the response functions to details of
the structure of the Roper resonance is investigated. The possibility of
disentangling this resonance from the contribution of 
$\Delta$ electroproduction in nuclei is addressed.
Finally, the contributions of the Roper resonance to the 
longitudinal scaling function and to the Coulomb sum rule are also explored.
\end{abstract}

\end{frontmatter}

\section{Introduction}
\label{intro}

Parity-conserving (PC)
inclusive electron scattering from nuclei has attracted
considerable attention as a probe of nuclear structure. 
Virtual photons penetrate deeply inside the nucleus allowing one to test
modeling of various aspects of the nuclear many-body problem including
effects from
nuclear correlations, meson exchange currents and relativistic
effects~\cite{Boffi:1993gs,Amaro:2002mj}. Parity-violating (PV) electron 
scattering from nuclei --- inclusive scattering of longitudinally
polarized electrons --- goes even further by opening 
the possibility of obtaining new information not
only on nuclear, but also on nucleon structure, 
especially on the strangeness content of the nucleon. 
Indeed, a basic motivation for studying PV asymmetries
in nuclei with polarized electrons is that scattering from a 
free proton alone is insufficient when attempting to disentangle 
all of nucleon's 
form factors and somewhere a neutron (hence, a nucleus) must be
involved~\cite{Donnelly:1992qy}.  

Many studies have been focused on the energy
region of the quasielastic (QE) peak and the $\Delta$
resonance~\cite{Amaro:2002mj,Gil:1997bm}. However, there
is increasing interest in the so called second resonance region,
namely where the $N^*(1440)P_{11}$ (Roper), $N^*(1520)D_{13}$
and $N^*(1535)S_{11}$ resonances are to be found. 
Understanding the properties of these $N-N^*$ 
transitions is a challenge for quark models and 
relates to basic issues such as the nature of quark-hadron duality and the
roles played by gluons in the baryon excitation spectrum. The experimental 
study of meson electroproduction on the proton, currently underway at
JLab, is presently providing data in this 
region~\cite{Burkert:2002nr,Burkert:2001bx}. 
Similar studies in nuclei would add information on 
the in-medium dynamics of these resonances and on possible
modifications of baryon properties inside the
nucleus~\cite{Lehr:1999zr}.  

The $N^*(1440)P_{11}$ is one of the most
puzzling of the resonances occurring just above the $\Delta$(1232). 
It is usually viewed as a radial excitation of a
three-quark nucleon state --- a breathing mode. However, the standard 
non-relativistic constituent quark
modeling fails to describe all of its properties, in particular its 
mass which is overestimated.
This outcome has prompted further studies, pointing to
the relevance of  chiral symmetry~\cite{Glozman:1996fu}, 
relativistic corrections~\cite{Dong:1999cz}, meson
clouds~\cite{Dong:1999cz,Cano:1998wz} or configuration mixing due to gluon
exchange~\cite{Cardarelli:1997vn}. 
Alternative approaches describe the Roper as 
a Skyrme soliton~\cite{Mattis:1985dh,Biedenharn:1985he}, 
a hybrid state with a
large gluonic component~\cite{Li:1992yb}, a nucleon-sigma 
molecule~\cite{Krehl:1999km} or 
as a chiral soliton within the chromodielectric model~\cite{Alberto:2001fy}. 
All of these models predict specific behaviors for the
electroproduction amplitudes $A^{p,n}_{1/2}$, $S^{p,n}_{1/2}$, which, in turn,
have consequences for the nuclear response functions. The experimental
information available so far comes from the multipole analysis of
Gerhardt~\cite{Gerhardt:1980yg} and is clearly insufficient to
allow a stringent test of these theories. It is also worth
mentioning the relevant role played by the $N^*(1440)$ in the
description of several hadronic reactions at intermediate energies,  
$\pi N \raw \pi \pi N$~\cite{Oset:1985wt}, 
$N N \raw N N \pi \pi$~\cite{Alvarez-Ruso:1998mx} and especially 
the $(\alpha, \alpha')$
reaction on a proton target where the isoscalar excitation of the Roper
clearly stands out in the data~\cite{Morsch:1992vj,Hirenzaki:1996js}.

The aim of this work is to make a first step towards a theoretical
description of the nuclear response functions (both parity-conserving and 
-violating) in the second resonance region. We calculate these 
responses for the Roper resonance excitation in nuclei using the 
simple (perhaps oversimplified, but at least respecting Lorentz and
gauge invariance)  Relativistic Fermi Gas (RFG)
model~\cite{Alberico:1988zx}. The paper is organized as follows: first
we consider the PC response and, after discussing the
general formalism, we calculate the RFG response functions in the domain of the
$N-N^*(1440)$ transition using for this form factors
related to the helicity amplitudes. We present results corresponding to
various parameterizations and model calculations of the Roper's structure
available in the literature, emphasizing their specific impacts on 
the nuclear response functions. Next we explore the consequences of including 
the Roper excitation on the scaling and {\it superscaling} properties of the
nuclear response functions and on the Coulomb sum rule. Finally, 
we perform a similar investigation in the PV case.   

\section{Parity-conserving response functions}
\la{PC}

\subsection{General formalism}
\la{gene}

We consider the scattering of an electron with initial and final
four-momenta $k= (E, \vk)$ and $k'= (E', \vk')$,
respectively, from a nucleus. 
The momentum transferred to the latter, namely 
$q = k-k'=(\omega, \vq)$, is carried by a single virtual photon. 
As in past work, the set of dimensionless variables
\be
\la{variaq}
\kappa=(\lambda, \bkappa)=
\left(\frac{\omega}{2m_{\chic{N}}}, \frac{\vq}{2 m_{\chic{N}}} \right)\,,\quad
\tau=\bkappa^2-\lambda^2=-\frac{q^2}{4 m^2_{\chic{N}}} = \frac{|Q^2|}{4
m^2_{\chic{N}}}  
\ee
\be
\la{varian}
\eta =(\epsilon, \boeta)=
\left(\sqrt{1+\frac{\vp^2}{m_{\chic{N}}^2}},\
\frac{\vp}{m_{\chic{N}}}\right)\,,
\ee
\be
\la{fermi}
\eta_{\chic{F}} = \frac{k_{\chic{F}}}{m_{\chic{N}}}\,,\quad
\epsilon_{\chic{F}}=\sqrt{1+\eta_{\chic{F}}^2}\,,\quad
\xi_{\chic{F}}=\epsilon_{\chic{F}}-1
\ee
($m_{\chic{N}}$ being the nucleon mass and $k_{\chic{F}}$ the Fermi
momentum) for the nucleon is used.

In the Born approximation and in the 
ultra-relativistic limit ($m_e \raw 0$), the inclusive 
differential cross section reads~\cite{Donnelly:1975ze}
\be
\la{diff}
\frac{d\sigma}{d\Omega' dE'} = \sigma_{\chic{M}}
\frac{1}{2 E E' \cos^2{\theta_e/2}} L_{\mu \nu} W^{\mu \nu}  
= \sigma_{\chic{M}} \left( v_{\chic{L}} R^{\chic{L}} +  v_{\chic{T}}
R^{\chic{T}} \right)\,, 
\ee
$\sigma_{\chic{M}}$ being the Mott cross section, $\theta_e$ the
electron scattering angle and $R^{{\chic{L}}({\chic{T}})}$ the
longitudinal (transverse) response functions. The kinematical factors 
$v_{{\chic{L}}({\chic{T}})}$ are 
\be
\la{vLT}
v_{\chic{L}}=\frac{\tau^2}{\bkappa^4}\,,\quad
v_{\chic{T}}= \frac{\tau}{2 \bkappa^2} +
\tan^2{\frac{\theta_e}{2}}
\ee
and the leptonic tensor $L_{\mu \nu}$ is
\be
\la{lepto}
L_{\mu \nu} =\frac{1}{4} \mathrm{Tr} \left[k\!\!\!/ \gamma_\mu
k'\!\!\!\!/ \gamma_\nu \right] = k_\mu k'_\nu + k'_\mu k_\nu - g_{\mu
\nu} k \cdot k'\,. 
\ee
The response functions are connected to the nuclear tensor 
according to the following relations
\be
\la{RLT}
R^{\chic{L}} = W^{00}\,,\quad R^{\chic{T}} = W^{11}+W^{22}\,.
\ee

In our case, $W^{\mu \nu}$  accounts for the 
$N-N^*(1440)$ transition inside the nucleus. Analogously to the case of
the $N-\Delta$ transition~\cite{Amaro:1999be}, it is given by
\be
\la{hadro}
W^{\mu \nu} = \int^{\mu^*_{max}}_{\mu^*_{min}} d\mu^* G(\mu^*)
W_0^{\mu \nu}(\mu^*)\,,
\ee
where $\mu^*$, the ratio between the Roper invariant
mass W and the nucleon mass, ranges from
$\mu^*_{min}=1+(m_\pi/m_{\chic{N}})$ to 
$\mu^*_{max}=\sqrt{(\epsilon_{\chic{F}}+2
\lambda)^2-(\eta_{\chic{F}}-2 |\bkappa|)^2}$. Moreover, 
we parametrize the spectral function of the Roper resonance in the
form
\be
\la{G}
G(\mu^*) = \frac{1}{2 \pi}
\frac{\Gamma(W)/m_{\chic{N}}}{\left(\mu^*-m^*/m_{\chic{N}}
\right)^2 + \Gamma^2(W)/(4 m^2_{\chic{N}})}\,,  
\ee
where $m^* = 1440$~MeV is the resonance mass and $\Gamma(W)$ its total 
width. Our specific expression for this energy-dependent total width is given 
in the Appendix. Finally, $W_0^{\mu \nu}(\mu^*)$ is the standard
RFG tensor associated with a {\it narrow} $N^*$ of mass $W = m_{\chic{N}}\mu^*$,
namely
\be
\la{hadrost}
W_0^{\mu \nu}(\mu^*)=\frac{3 {\mathcal N}}{4 |\bkappa| m_{\chic{N}}
\eta_F^3} \int_{\epsilon_0}^{\epsilon_F} f^{\mu \nu}(\epsilon,\mu^*) d\epsilon
\,,
\ee
where 
\be
\la{e0}
\epsilon_0=|\bkappa| \sqrt{1/\tau +\rho^2}-\lambda \rho
\ee
is the minimum energy of the struck nucleon as a function of
$|\bkappa|$, $\lambda$ and the inelasticity parameter $\rho$, 
which is defined as~\cite{Amaro:1999be}
\be
\la{inel}
\rho = 1+\frac{1}{4 \tau}(\mu^{*2}-1)\,.
\ee
The particle number ${\mathcal N}$ corresponds to the proton and neutron
number, respectively. The nuclear responses are clearly obtained by adding 
the contributions from the two species, with weightings ${\mathcal
  N}=Z$ and ${\mathcal N}=N$, respectively.

Concerning the single-nucleon tensor $f_{\mu \nu}$, its standard form 
is given by~\cite{Donnelly:1992qy}
\be
\la{gener}
f^{\mu \nu} =
 -w_1(\tau, \mu^*)\left(g^{\mu \nu}+\frac{\kappa^\mu \kappa^\nu}{\tau}\right)
             +w_2(\tau, \mu^*) V^\mu V^\nu \,,
\ee
where $V_\mu = \eta_\mu+\kappa_\mu \rho$. The insertion of
Eq.~(\ref{gener}) into Eq.~(\ref{hadrost}) leads then to the following
expressions for the longitudinal and transverse response functions 
\be
\la{ave}
R^{{\chic{L}}({\chic{T}})} (|\vq |, \omega) =   
\int^{\mu^*_{max}}_{\mu^*_{min}} d\mu^* G(\mu^*) 
R_0^{{\chic{L}}({\chic{T}})}(|\vq |, \omega, \mu^*) \,,   
\ee
where 
\be
\la{R0LT}
R_0^{{\chic{L}}({\chic{T}})}(|\vq |, \omega, \mu^*) = 
\frac{3 {\mathcal N}}{4 |\bkappa| m_{\chic{N}} \eta_{\chic{F}}^3}
\xi_{\chic{F}}\, \Theta (1 - \psi^{*2}) (1 - \psi^{*2})  
U^{\chic{L}(\chic{T})}(|\vq |,\omega, \mu^*) 
\ee
with 
\bea
\la{UL}
U^{\chic{L}} &=& \frac{\bkappa^2}{\tau} 
\left[ (1+\tau \rho^2) w_2(\tau,\mu^*)-w_1(\tau,\mu^*)
+w_2(\tau,\mu^*){\mathcal D}(\kappa, \mu^*)
\right]\,, \\[0.2cm]
\la{UT}
U^{\chic{T}} &=& 2 w_1(\tau,\mu^*) +w_2(\tau,\mu^*){\mathcal
D}(\kappa,\mu^*)\,, 
\eea
\bea
\la{D}
{\mathcal D} (\kappa,\mu^*) &=&  \frac{\tau}{\bkappa^2} \left[ 
(\lambda\rho+1)^2+(\lambda\rho+1)(1+
\psi^{*2})\xi_{\chic{F}}+\frac{1}{3}(1
+\psi^{*2}+\psi^{*4})\xi_{\chic{F}}^2\right] 
\nonumber \\[0.1cm]
&-&(1+\tau\rho^2)\,.
\eea
Moreover, 
\be
\la{scal}
\psi^{*2} = \frac{\epsilon_0 -1}{\xi_{\chic{F}}} \,
\ee
is the squared scaling variable associated with the resonance.
The above expressions are independent of the specific transition under
consideration: they can be used to describe the $\Delta$ region 
(after replacing $m^*$ by $m_\Delta$ in Eq.~(\ref{G}) and modifying 
$\Gamma(W)$ accordingly) 
as well as the quasielastic peak (in the $\Gamma \raw 0$ limit setting
$m^*=m_{\chic{N}}$)~\cite{Donnelly:1992qy,Amaro:1999be,Alvarez-Ruso:2000bx}.
In other words, all of the information about the $N-N^*(1440)$ transition
is embedded in the functions $w_{1,2}(\tau,\mu^*)$. In order to relate
these to the transition form factors (and to the helicity amplitudes), we
need to write $f^{\mu \nu}$ in terms of the $N-N^*(1440)$
electromagnetic current. 

\subsection{Hadronic tensor and form factors}

The $N-N^*(1440)$ tensor reads
\be
\la{hadtensor}
f^{\mu \nu} = \frac{1}{2} \mu^* \mathrm{Tr}
\left[\frac{(p\!\!\!\!\:/+m_{\chic{N}})}{2 m_{\chic{N}}} \left(\gamma_0 
J^{\dagger \mu} \gamma_0 \right) \frac{(p\!\!\!\!\:/'+W)}{2 W} J^\nu 
\right] \,,
\ee
where the spinor matrix element of the current, written in terms 
of form factors and gauge invariant 
operators~\cite{Devenish:1976jd,Weber:1990fv,Alvarez-Ruso:1998jr}, is given by
\be
\la{N*curr}
J^\alpha = \bar u_{N^*}(p') \left[ F_1(q^2)\left( q \!\!\!\!\: /\, 
q^{\alpha} - 
q^2 \gamma^{\alpha}\right) + i F_2(q^2) \sigma^{\alpha \beta} q_{\beta}   
\right] u(p)\,.
\ee
Notice the structure of the current which is very similar to its
nucleonic counterpart (but for obvious redefinitions of the form factors)
except for the $q\!\!\!\!\,/\, q^{\alpha}$ part. For the
nucleon, the form factor associated with this operator has to vanish
to ensure current conservation, but not for the Roper, since the 
mass of the latter differs from that of the nucleon.

Substituting Eq.~(\ref{N*curr}) into Eq.~(\ref{hadtensor}) and casting the
result in the form of Eq.~(\ref{gener}), one gets $w_{1,2}$ in terms
of the form factors according to
\bea
\la{w1}
w_1 &=& \D   \left[ \tau +\frac{(\mu^* -1)^2}{4} \right] G_{\chic{M}}^2
\,, \\[0.2cm]
\la{w2}
w_2 &=& \D \frac{1}{\D 1 + \frac{4 \tau}{(\mu^* +1)^2}} 
\left[ G_{\chic{E}}^2 + \frac{4 \tau}{(\mu^* +1)^2} G_{\chic{M}}^2 \right]\,,
\eea
where $G_{\chic{E},\chic{M}}$ are defined analogously to the
Sachs form factors of the
nucleon~\cite{Cardarelli:1997vn,Weber:1990fv}, 
{\it i.e.,}
\bea
\la{GE}
G_{\chic{E}} &=&\D 4 m_{\chic{N}}^2 \tau \left[ F_1 -
\frac{F_2}{m_{\chic{N}} (\mu^* + 1)} \right] \,, \\[0.2cm]
\la{GM}
G_{\chic{M}} &=&\D 4 m_{\chic{N}}^2 \tau F_1 + m_{\chic{N}} (\mu^* +1)
F_2\,. 
\eea
Indeed, when $\mu^* \raw 1$, Eqs.~(\ref{w1},\ref{w2}) coincide with the
well-known expressions for the nucleon (see for instance~\cite{Donnelly:1992qy}). 

\subsection{Helicity amplitudes}

Electromagnetic transitions between a nucleon and a resonant state are
often expressed as helicity amplitudes, which describe transitions
between a nucleon state with helicity $\lambda = 1/2$ and a resonant
state with $\lambda' = 1/2$ or $3/2$. In the case of the Roper two
helicity amplitudes $A_{1/2}$ and $S_{1/2}$ should be introduced. They are
defined as~\cite{Weber:1990fv}
\begin{eqnarray}
\la{Adef}
A_{1/2}^{p(n)}(Q^2) & = & \sqrt{\frac{2 \pi \alpha}{k_R}} {\Bra{N^*
\downarrow}} \epsilon^{\chic (+)}_\mu 
 J^{p(n)\mu} {\Ket{N \uparrow}}\,,\\
\la{Sdef}
S_{1/2}^{p(n)}(Q^2) & = & \sqrt{\frac{2 \pi \alpha}{k_R}}
\frac{|{\mathbf{q}|}}{Q^2}
{\Bra{N^*\uparrow}} \epsilon^{\chic (0)}_\mu 
 J^{p(n)\mu} {\Ket{N \uparrow}}\,,
\nonumber
\end{eqnarray}
where $\alpha$ is the electromagnetic fine-structure constant, 
$k_R$ is the energy of a real
photon equivalent to the virtual one and 
$\epsilon^{\chic (+,0)}$ stand for the transverse and
longitudinal polarizations of the virtual photon. Inserting
Eq.~(\ref{N*curr}) in the above expressions and using the definitions 
of $G_{\chic E,M}$ given in Eqs.~(\ref{GE}, \ref{GM}), it is
straightforward to relate the helicity amplitudes to the form
factors~\cite{Cardarelli:1997vn,Alvarez-Ruso:1998jr} according to
\bea
\la{A}
A^{p(n)}_{1/2}(\tau ) &=& f(\tau,\mu^*) G^{p(n)}_{\chic{M}}
(\tau,\mu^*)   \,, \\[0.2cm] 
\la{S}
S^{p(n)}_{1/2}(\tau ) &=& f(\tau,\mu^*) \frac{\sqrt{[(\mu^*-1)^2+4 \tau]
[(\mu^*+1)^2+4 \tau]}}{8 \sqrt{2} \tau}\, \times \nonumber \\
&& \frac{\mu^* + 1}{\mu^*} G^{p(n)}_{\chic{E}} (\tau,\mu^*) \,, 
\eea
with 
\be
\la{f}
f(\tau,\mu^*)= \sqrt{ \frac{2 \pi \alpha}{m_{\chic{N}}}
\frac{(\mu^*-1)^2+4 \tau}{\mu^{*2} -1}} \,.  
\ee

\begin{figure}[h!]
\begin{center}
\includegraphics[width=0.5\textwidth]{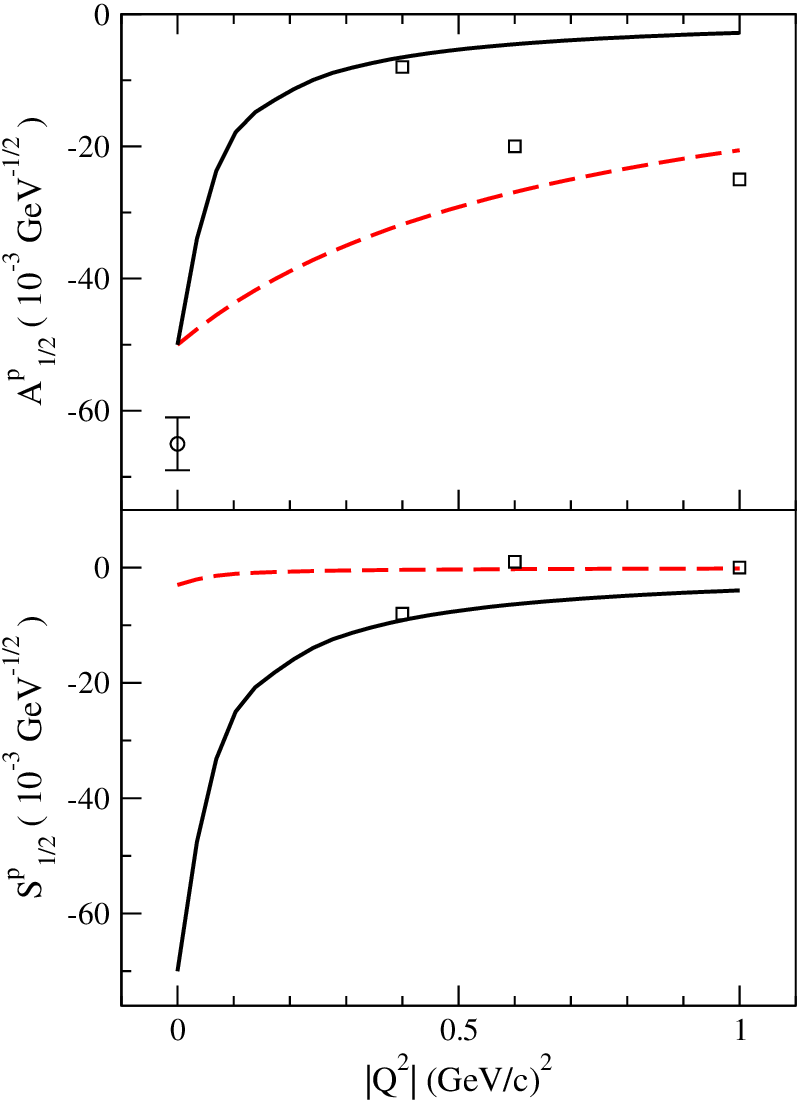} 
\caption{$N-N^*(1440)$ helicity amplitudes as a function of $|Q^2|$ for
two different fits, Gh1 (solid lines) and Gh2 (dashed
lines), of the data~\cite{Gerhardt:1980yg,Li:1992yb}. The squares are
the results of Gerhardt's  analysis at fixed energies and the circle is
the value at the photon point~\cite{Groom:2000in}.} 
\label{fig1}
\end{center}
\end{figure} 
As already mentioned, 
the only well established experimental information about these
amplitudes was obtained by Gerhardt from a partial wave analysis of
the data taken at NINA  and DESY~\cite{Gerhardt:1980yg}. A fit in the
second resonance region was performed 
independently at three fixed values of $Q^2$ and in the whole
range (0-1~(GeV$/c$)$^2$) for all of the NINA and DESY data separately. 
The results, 
converted into helicity amplitudes from the original 
$M_{1-}$ and $S_{1-}$ multipoles by Li {\it et al.}~\cite{Li:1992yb},
are shown in Fig.~\ref{fig1}. These amplitudes are for the $p-p^*(1440)$
transition; there is no information on the $Q^2$-dependence of the 
neutron form factors and only the value at the photon point ($Q^2 = 0$) is
known~\cite{Groom:2000in}. 
One should also bear in mind that the
analysis includes model-dependent assumptions; the systematic error is
estimated to be no smaller than $\pm 12\times
10^{-3}$~GeV$^{-1/2}$~\cite{Li:1992yb}. On the positive side, the preliminary
result of an analysis of the pion electroproduction data measured in the
CLAS detector at JLab at the $|Q^2| = 0.4$~(GeV$/c$)$^2$ 
is in excellent agreement with Gerhardt's value~\cite{Burkert:2002nr}.    

\begin{figure}[h!]
\begin{center}
\includegraphics[width=.995\textwidth]{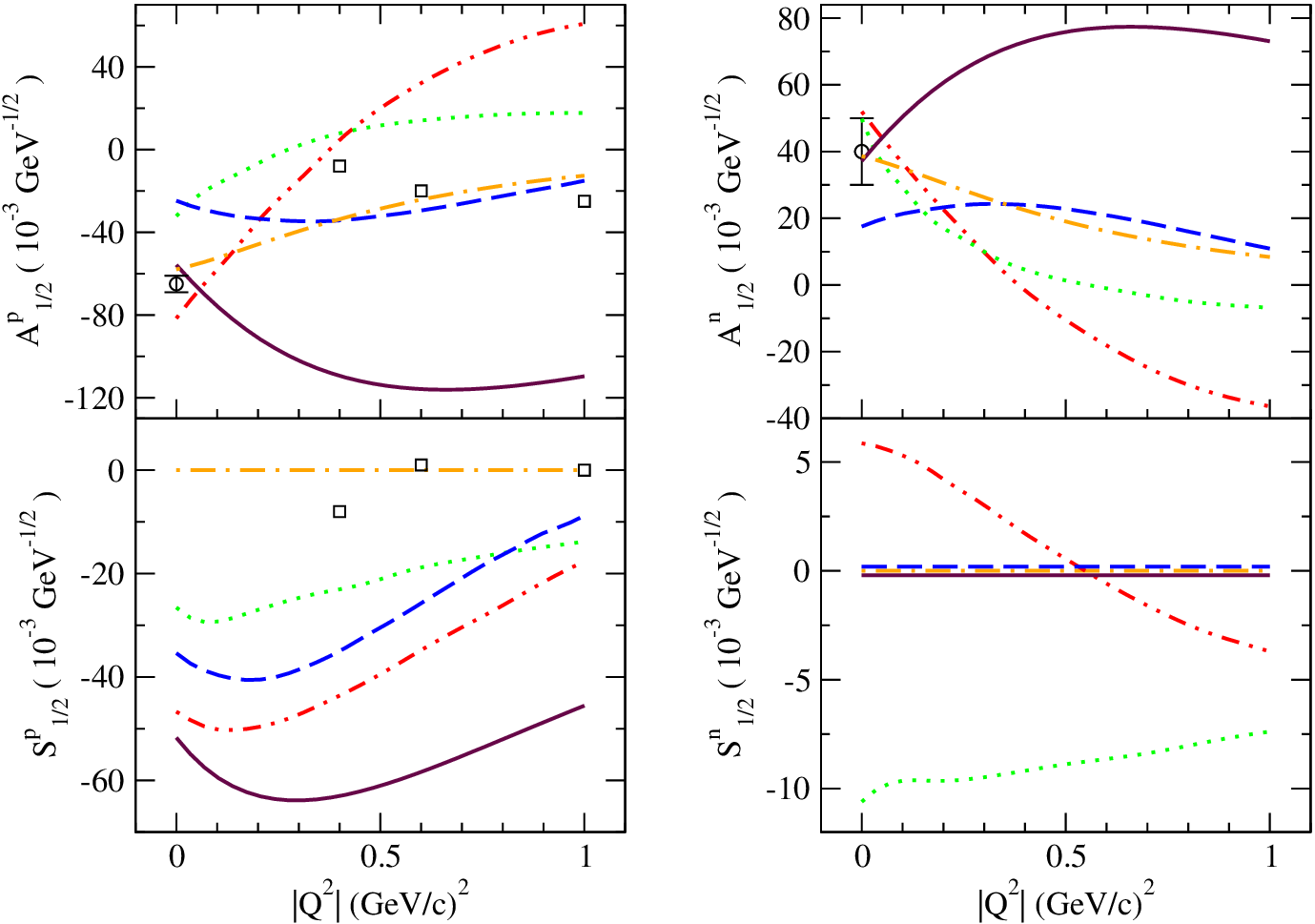}
\caption{Transverse ($A_{1/2}$) and longitudinal ($S_{1/2}$)
electroproduction amplitudes for the $N^*(1440)$ calculated with
various models: NRQM (solid line)~\cite{Li:1992yb}, 
hybrid model (dash-dotted line)~\cite{Li:1992yb}, 
LF (dotted line)~\cite{Cardarelli:1997vn}, 
ChD (dashed line)~\cite{Alberto:2001fy} 
and EVMD (dash-double-dotted line)~\cite{Cano:1998wz}.}
\label{fig2}
\end{center}
\end{figure}

The $N-N^*(1440)$ transition amplitudes have been studied using 
various models with a wide diversity of results. Some of these are shown
in Fig.~\ref{fig2}, namely, the prediction
from the non-relativistic quark model (NRQM)~\cite{Li:1992yb}, the hybrid
model~\cite{Li:1992yb}, the light-front relativistic quark model (LF)
calculation of~\cite{Cardarelli:1997vn}, the chiral
chromodielectric (ChD) model (self-consistent
calculation)~\cite{Alberto:2001fy} and the extended vector-meson
dominance (EVMD) model of~\cite{Cano:1998wz} ($V_I$ potential). Notice that
the hybrid model predicts $S^p_{1/2}=0$, whereas $S^n_{1/2}=0$ is obtained
for both the NRQM and hybrid model. The chromodielectric model also 
yields an $S^n_{1/2}$ consistent with zero. 

\subsection{Results and discussion}
\la{rad}

\begin{figure}[th!]
\begin{center}
\vspace{1.cm}
\includegraphics[width=\textwidth]{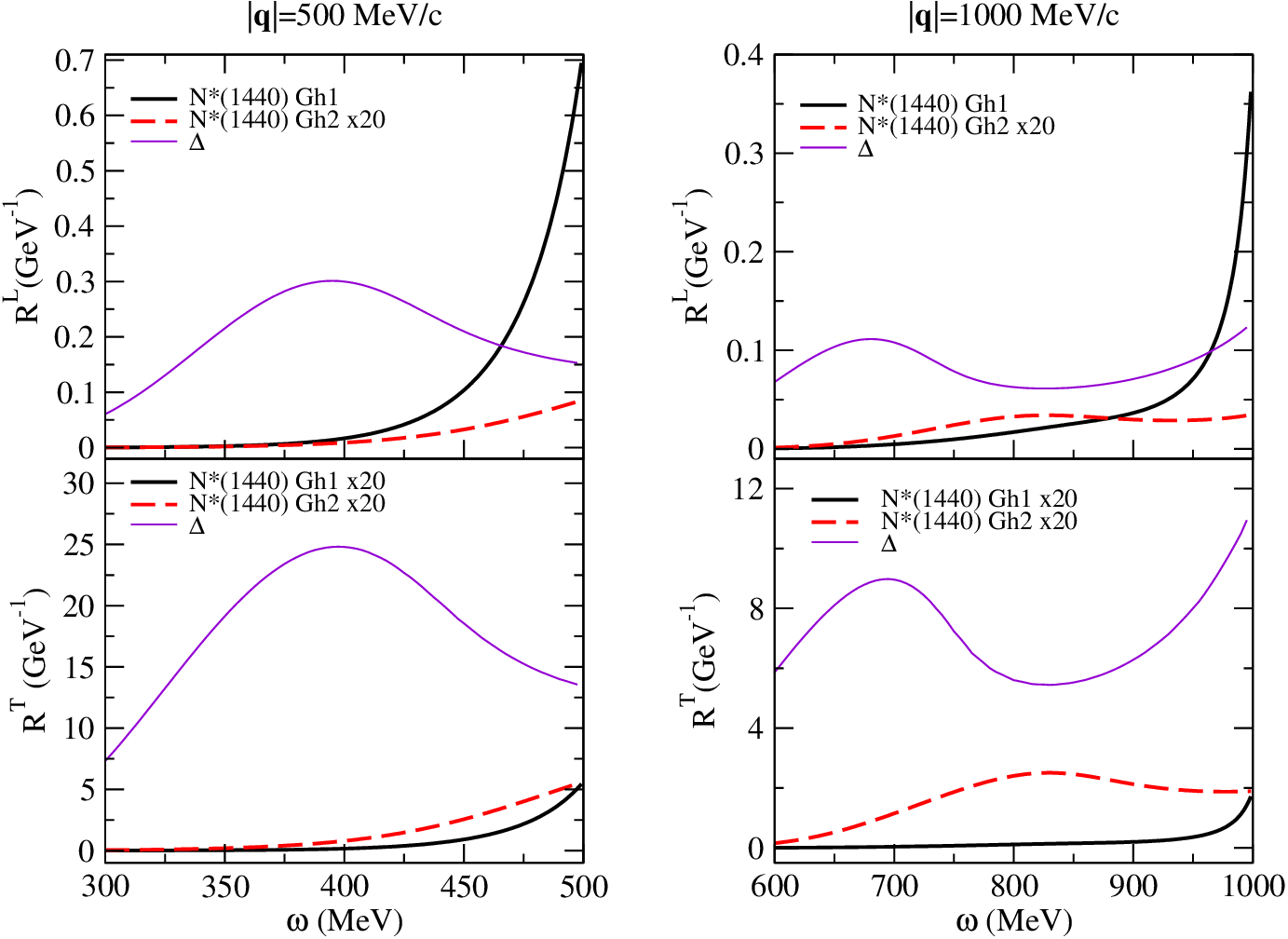}
\caption{PC $N-N^*(1440)$ response functions for $^{12}$C at two
values of the transferred 3-momentum and for two different
parameterizations of the helicity amplitudes, Gh1 and Gh2 
(see Fig.~\ref{fig2}). The $N-\Delta$ responses are also shown. Notice
that some curves have been multiplied by a factor 20 to make them
visible. }
\label{fig3}
\end{center}
\end{figure}

The RFG PC  $N-N^*(1440)$ response functions are shown in
Fig.~\ref{fig3} for two values of the transferred 3-momentum
using the two empirical parameterizations for the helicity amplitudes
displayed in Fig.~\ref{fig1}. Since
there are no measurements on the neutron, we assume the NRQM relations
\be
\la{pn}
A^n_{1/2}=-\frac{2}{3} A^p_{1/2}\,,\quad S^n_{1/2}=0\,. 
\ee
The choice of
$k_{\chic{F}}=225$~MeV$/c$ and $Z=N=6$ corresponds to $^{12}$C. The
contribution of the $\Delta$, as calculated in~\cite{Amaro:1999be}, has 
also been included. The total RFG response will thus arise, in first
approximation, from the sum of the Roper and the $\Delta$ contributions.
In the case of the transverse response, the Roper's effect
is negligible compared with that of the $\Delta$ at all
energies. This does not come as surprise, since the
strong $N - \Delta$ $M1$ transition appears at full strength in this
observable. However, in the longitudinal channel the situation is
different: the leading terms
cancel~\cite{Amaro:1999be,Alvarez-Ruso:2000bx}, which results
in a longitudinal $\Delta$ response an order of magnitude smaller than the
transverse one, giving a chance for other mechanisms to show up. Indeed, 
in the upper panels of Fig.~\ref{fig3}, the Roper contribution
appears as a moderately sharp peak at the edges of the spectra. Its
size depends strongly on the parameterization of the helicity
amplitudes.  However, it should be kept in mind that near 
the light-cone the kinematical
factor $v_{\chic{L}}$ severely quenches the longitudinal channel, and 
hence it is doubtful if the Roper can be detected in this domain.
\begin{figure}[h!]
\vspace{1.3cm}
\begin{center}
\includegraphics[width=\textwidth]{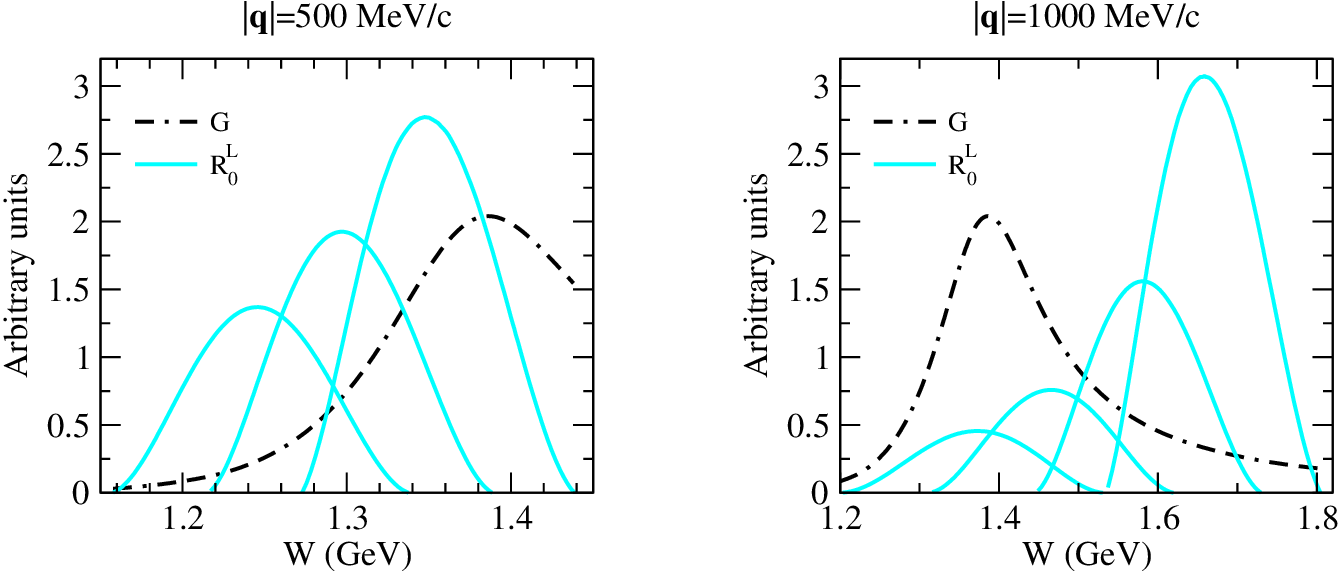}
\caption{Combined plot of $G$ and $R_0^{\chic{L}}$ functions as a
function of the $N^*$ invariant mass. $R_0^{\chic{L}}$ is shown at
various $\omega$'s which are, from left to right, 400, 450 and
500~MeV on the left panel and 750, 830, 930 and 1000~MeV on the right
panel. }
\label{fig4}
\end{center}
\end{figure}

Be it as it may, in order to understand the behavior of the Roper response near
the light-cone, we have plotted the spectral function $G$ [see
Eq.~(\ref{G})] and
$R_0^{\chic{L}}$ as a function of the $N^*$ invariant mass in
Fig.~\ref{fig4}. Their
convolution for a given $\omega$ value yields the response according to
Eq.~(\ref{ave}). Here we have used the parameterization
Gh2 (dashed lines in Figs.~\ref{fig1} and \ref{fig3}). At larger values of
$\omega$, $R_0^{\chic{L}}$ probes higher invariant masses --- hence
its value at the maximum increases. 
At $|\vq |=500$~MeV$/c$, the regions of higher 
$R_0^{\chic{L}}$ correspond to larger $G$ values, producing a
monotonic rise of $R^{\chic{L}}$ as seen in the upper left panel of
Fig.~\ref{fig3}. The situation is completely different at
$|\vq |=1000$~MeV$/c$ where small  $R_0^{\chic{L}}$ and large $G$ are
combined to produce a local maximum around $\omega=830$~MeV (see the
dashed line in the upper right panel of Fig.~\ref{fig3}) followed by
a very shallow minimum around $930$~MeV; then, the rapid growth of
$R_0^{\chic{L}}$ compensates the decrease of $G$ at the edge of the
spectrum.   

Further information can be extracted from
Fig.~\ref{fig4}. At $|\vq |=1000$~MeV$/c$ and for large values of $\omega$,
where the Roper contribution might be significant, the relevant values of $W$
are well above the Roper peak. Therefore the
response functions will not be dominated by the $\Delta$ and the Roper
alone. Indeed many heavier resonances will contribute, making 
unreliable the information about the $N-N^*(1440)$ transition obtained on the
basis of the data at this value of $|\vq |$. However, at $|\vq |=500$~MeV$/c$, the
relevant values of $W$ are below $1.44$~GeV. It is then plausible to
assume that the contribution of resonances heavier than the Roper is
small, so that, relying on our present knowledge of the $N-\Delta$
transition, useful information about the Roper can eventually be
obtained.

It is instructive to take a closer look at the responses in the light-cone 
limit ($\omega \raw |\vq |$, {\it i.e.,} $\tau \raw 0$; note that this
limit is actually unobtainable for inelastic electron scattering even
when the scattering angle approaches zero). A straightforward
calculation yields
\bea
\la{UL0}
U^{\chic{L}} (\lambda \raw |\bkappa |) &=&\left( \frac{m_{\chic{N}}}{2
\pi \alpha} \right) \left[ 
8 \bkappa^2 \frac{\mu^{*2}}{\mu^{*2}-1} 
S_{1/2}^2(0) + |\bkappa | \xi_{\chic{F}} (1-\psi_0^2)
A_{1/2}^2(0)\right] \,, \\[0.2cm]
\la{UT0}
U^{\chic{T}} (\lambda \raw |\bkappa |) &=& \left( \frac{m_{\chic{N}}}{2
\pi \alpha} \right) \frac{1}{2} (\mu^{*2} - 1)
A_{1/2}^2(0)\,,
\eea
with 
\be
\la{psi0}
\psi_0^2 = \frac{(\mu^{*2}-1-4 |\bkappa |)^2}{8 \xi_{\chic{F}}
(\mu^{*2} - 1) |\bkappa |}\,.
\ee
The insertion of the relevant numbers in Eq.~(\ref{UL0}) reveals the
fact that the 
factor in front of $A_{1/2}^2(0)$ is about two orders of magnitude smaller than
the one in front of $S_{1/2}^2(0)$ (notice the smallness of
$\xi_{\chic{F}}=0.03$). Therefore, the peak observed at
the edge of the longitudinal response function is completely
dominated by the $S_{1/2}$ amplitude evaluated close to the origin. It is then
obvious that the different values for $S^p_{1/2}$ close to the
origin predicted by Gh1 and Gh2 fits (lower panel in Fig.~\ref{fig1}) are 
responsible for the large differences in the corresponding longitudinal response
observed in Fig.~\ref{fig3}. By the same argument, Eq.~(\ref{UT0}) and the
upper panel of Fig.~\ref{fig1} explain why the transverse responses using 
the two parameterizations coincide as  
$\omega \raw |\vq |$, the curve obtained with 
Gh1 being above the other in the vicinity of this point.                        

Finally, we present the electromagnetic responses at $|\vq |=500$~MeV$/c$
calculated with the helicity amplitudes obtained from the various models
shown in Fig.~\ref{fig2}. As expected from the above discussion, the
size of the peak of the longitudinal response is determined by the strength of 
$S_{1/2}$ at $Q^2 \raw 0$. There is a big gap between the large result 
obtained with the
NRQM and the almost negligible one found for the hybrid model where
$S^p_{1/2}=S^n_{1/2}=0$. Notice also that the pattern is completely
different in the case of $R^{\chic{T}}$.     

\begin{figure}[h!]
\vspace{.4cm}
\begin{center}
\includegraphics[width=0.55\textwidth]{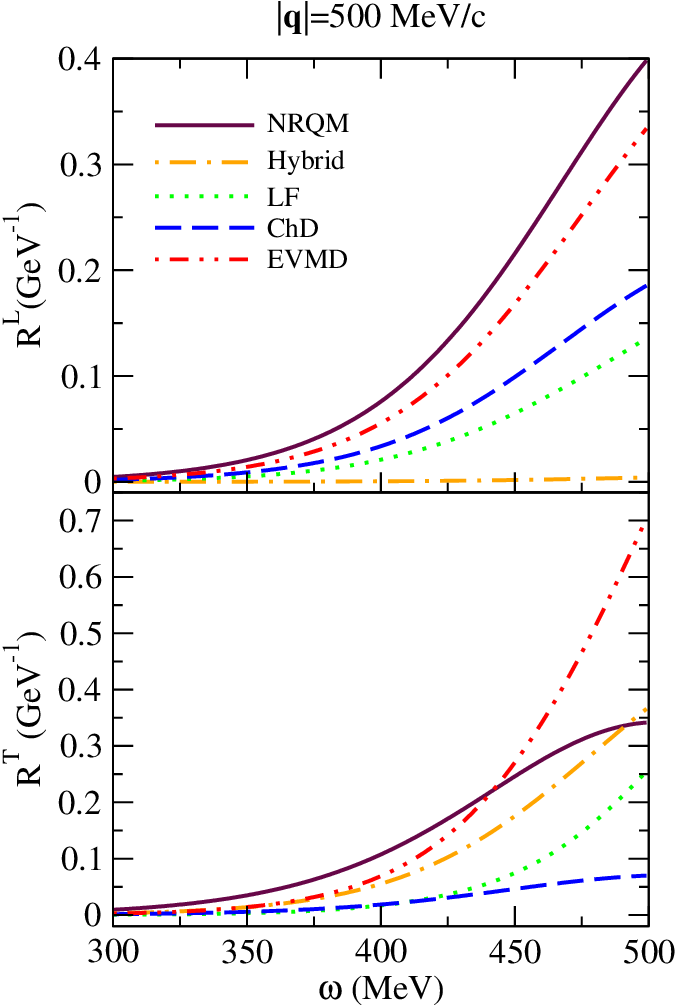}
\caption{Electromagnetic $N-N^*(1440)$ responses at $|\vq |
=500$~MeV$/c$ calculated using the helicity amplitudes from the models
displayed in Fig.~\ref{fig2}.}
\label{fig5}
\end{center}
\end{figure}

\section{Scaling behavior of the Roper responses}
\la{Scaling}

Study of the scaling properties of the cross sections and response
functions in inclusive electron scattering provides important
information about what constitute the relevant 
reaction mechanisms in the nucleus~\cite{Barbaro:1998gu,Maieron:2001it}. 
This access to scaling is achieved by dividing
the response functions by a function such that the physics related to the 
quasifree (elastic) scattering process on a single (moving) 
nucleon is removed. In this way, 
contributions from other mechanisms such as resonance excitation and meson
exchange currents, which in general do not scale, can be revealed.

Our aim in this section is to investigate the scale-breaking effects 
associated with the excitation of the Roper resonance. For this purpose, 
we have computed the scaling functions  
\be
\label{fLT}
f_{\chic{L(T)}}=k_{\chic{F}} \frac{R_{\chic{L(T)}}}{G_{\chic{L(T)}}}
\ee
as a function of the QE scaling variable
$\psi$~\cite{Donnelly:1992qy}, which can be easily obtained from $\psi^*$ of
Eqs.~(\ref{scal},\ref{e0}) by taking $\rho \raw 1$. The response functions 
$R_{\chic{L(T)}}$  are given by the sum of contributions from the 
quasi-elastic peak (taken from~\cite{Donnelly:1992qy})  
and the excitation of the $\Delta$ and Roper resonances. 
The functions $G_{\chic{L(T)}}$ correspond essentially to the
single-nucleon content of the nuclear responses. We take them as follows 
\bea
\label{GL}
G_{\chic{L}} &=& \frac{|\bkappa |}{2 \tau} \left[ Z G^2_{\chic{Ep}} + N
G^2_{\chic{En}} \right] \\[0.2cm]
\label{GT}
G_{\chic{T}} &=& \frac{\tau}{|\bkappa |} \left[ Z G^2_{\chic{Mp}} + N
G^2_{\chic{Mn}} \right] \,.
\eea
This choice differs slightly from the one employed in~\cite{Maieron:2001it}
(Eqs.~16-19) in that the (small) medium corrections included there are 
disregarded here.

\begin{figure}[th!]
\begin{center}
\includegraphics[width=\textwidth]{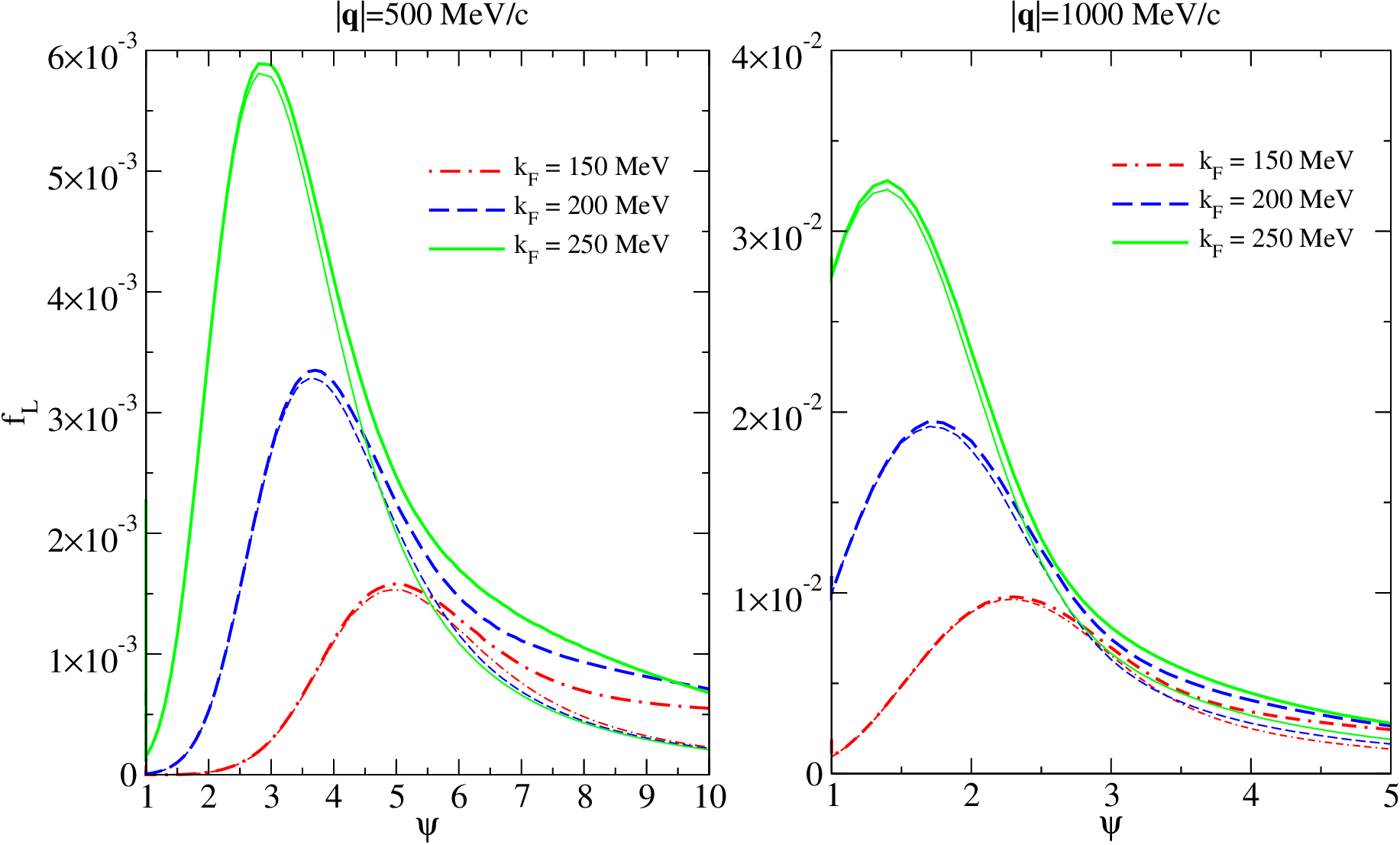} 
\caption{The longitudinal scaling function in resonance region, 
including the contributions 
of $\Delta$ alone (thin lines) and of the $\Delta$+Roper (thick lines)
for $|\vq|=500$ and 1000 MeV$/c$ and three values of $k_{\chic{F}}$. 
The $N^*$ contribution is computed using the Gh1 parameterization.}
\label{figsc1}
\end{center}
\end{figure}

In Fig.~\ref{figsc1} we display our results for 
$f_{\chic{L}}$ at $|\vq|=500$ and $1000$ MeV$/c$ and $\psi > 1$, {\it i.e.,} above the
quasielastic region where resonance excitation is important.  
As expected, violations of both first- and second-kind
scaling appear to set in. Above the
$\Delta$ peak, the Roper contribution to the scaling violation in the
longitudinal channel becomes important compared with that from the
$\Delta$, at least 
in the case when the Gh1 parameterization is used, although the overall size 
of the effect is tiny (we recall that $f_{\chic{L}}\simeq 0.7$ at $\psi=0$). 
Our findings of small scaling violation effects in the longitudinal channel 
is consistent with the experimental indications that $f_{\chic{L}}$ exhibits
superscaling behavior even above the quasielastic
peak~\cite{Donnelly:1999sw,Maieron:2001it}.  The scaling violation is much
more important in the transverse channel, but there the impact of the Roper
is negligible compared with that of the $\Delta$. 

\section{Resonance contribution to the Coulomb sum rule}
\la{CSR}

In this section, we investigate the contribution of the low-lying baryonic
resonances $\Delta$ and $N^*(1440)$ to the integrated longitudinal response
function, {\it i.e.} to the Coulomb sum rule. For this purpose, we compute the
following ratio 
\be
\la{ratio}
\mathcal{R} (|\vq |) = \frac{\D \int_0^{|\vq |} d\omega \left[
R^{{\chic{L}}}_{QE} (|\vq |, \omega) + R^{{\chic{L}}}_{\Delta, N^*} 
(|\vq |, \omega)
\right] }{\D \int_0^{|\vq |} d\omega R^{{\chic{L}}}_{QE} (|\vq |, \omega)} \,.
\ee
The integral over the energy transfer is restricted to the 
space-like region, the one explored 
in electron scattering experiments~\cite{Donnelly:1989dd}. 

\begin{figure}[h!]
\begin{center}
\vspace{0.1cm}
\includegraphics[width=0.6\textwidth]{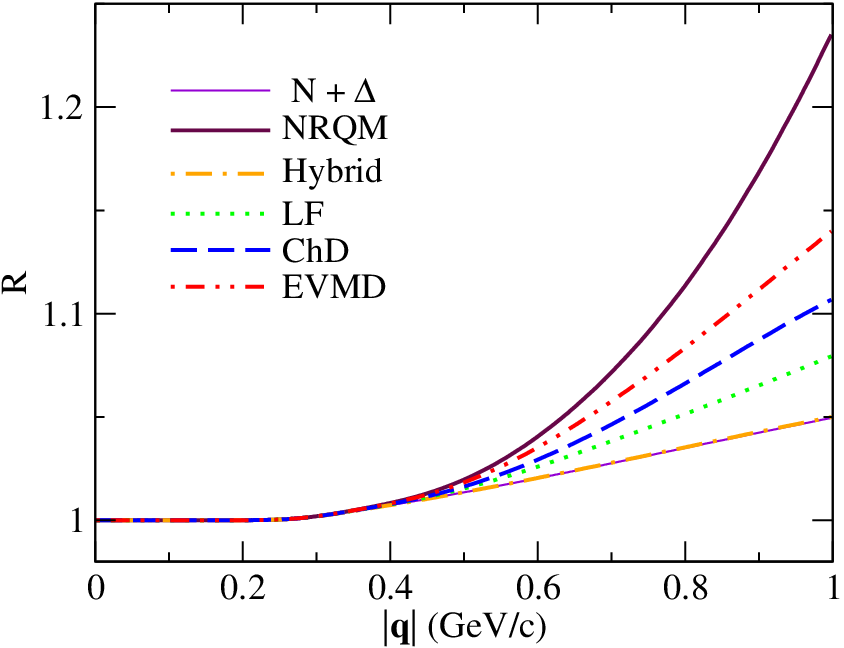} 
\caption{Resonance contribution to the Coulomb sum rule. The ratio 
$\mathcal{R}$ defined in Eq.~(\ref{ratio}) is evaluated including only the
$\Delta$ (solid thin line) and including both $\Delta$ and Roper
(thick lines) for the various models introduced above. Notice that the Roper
contribution in the hybrid model is negligible and practically coincides
with that of the $\Delta$ alone.}
\label{figcsr}
\end{center}
\end{figure}

The results are presented in Fig.~\ref{figcsr} using the $N - N^*(1440)$
transition amplitudes from the various models shown in Fig.~\ref{fig2}. 
The resonances are irrelevant below $|\vq | = 400$~MeV$/c$, but become
increasingly important at higher momentum transfers. 
The $\Delta$ accounts for at most 5\% whereas the impact of the Roper
is model-dependent and ranges from zero (Hybrid model) to almost 20\% at 
$|\vq | = 1$~GeV$/c$ for the NRQM.     

These findings imply that the resonances studied here should be accounted for when
deriving a realistic Coulomb sum rule, especially when this is used to extract 
$NN$ correlations from experiment, at momentum transfers above 500~MeV$/c$. 
By the same token the contribution from other $N^*$ resonances 
(beyond the Roper) is also likely to be important and should be studied.

\section{Parity-violating response functions}
\la{PV}

\subsection{General formalism}

Parity-violating effects due to weak interactions can be accessed
in polarized inclusive electron scattering experiments by measuring the helicity
asymmetry~\cite{Walecka:1977us} 
\be
\la{asim}
{\mathcal A}=\left( \frac{d\sigma}{d\Omega' dE'} \right)^{(PV)}
\Bigg/  \left( \frac{d\sigma}{d\Omega' dE'} \right)^{(PC)} \,.
\ee
The PC cross section is given by Eq.~(\ref{diff}) while
the PV one, defined as the difference of the cross
sections with opposite helicities of the initial electron, is 
expressed in terms of the corresponding PV response functions according to
\bea
\la{diffPV}
\left( \frac{d\sigma}{d\Omega' dE'} \right)^{(PV)} &\equiv&
\frac{1}{2} \left( \frac{d\sigma^+}{d\Omega' dE'} - 
\frac{d\sigma^-}{d\Omega' dE'} \right) \nonumber \\[0.1cm]
&=&  \sigma_{\chic{M}}
\frac{1}{2 E E' \cos^2{\theta_e/2}} \mathcal{A}_0 \widetilde{L}_{\mu \nu}
\widetilde{W}^{\mu \nu}  \nonumber \\[0.1cm]
&=& \sigma_{\chic{M}} \mathcal{A}_0 
\left( v_{\chic{L}} R^{\chic{L}}_{\chic{AV}} +  
v_{\chic{T}} R^{\chic{T}}_{\chic{AV}} + v_{\chic{T'}}
R^{\chic{T'}}_{\chic{VA}}   \right) \,, 
\eea 
where
\be
\la{A0}
\mathcal{A}_0 = \frac{G_{\chic{F}} |Q^2|}{2 \sqrt{2} \pi \alpha} \,,
\ee
$G_{\chic{F}}$ being the Fermi constant and 
\be
\la{vtp}
v_{\chic{T'}}=\sqrt{\frac{\tau}{\bkappa^2} +
\tan^2{\frac{\theta_e}{2}}} \tan{\frac{\theta_e}{2}}\,. 
\ee
The leptonic tensor is 
\be
\la{leptoPV}
\widetilde{L}_{\mu \nu} = a_{\chic{A}} L_{\mu \nu} + a_{\chic{V}}
\left(- i \epsilon_{\mu \nu \alpha \beta} k^\alpha k^{' \beta} \right)
\,, 
\ee
with $a_{\chic{A}}=-1$ and $a_{\chic{V}} =-(1-4
\sin^2{\theta_{\chic{W}}})$, $\theta_{\chic{W}}$ being the weak angle of
the standard electroweak theory. The following relations between the
hadronic PV
tensor $\widetilde{W}^{\mu \nu}$ and the response functions follow from 
Eq.~(\ref{diffPV}) 
\bea 
R^{\chic{L}}_{\chic{AV}} &\equiv& a_{\chic{A}}
\widetilde{R}^{\chic{L}} =
a_{\chic{A}}  \widetilde{W}^{00} \,,\\[0.2cm]
R^{\chic{T}}_{\chic{AV}} &\equiv& a_{\chic{A}} \widetilde{R}^{\chic{T}}  
= a_{\chic{A}} \left(
\widetilde{W}^{11} + \widetilde{W}^{22} \right) \,,\\[0.2cm] 
R^{\chic{T'}}_{\chic{AV}} &\equiv& a_{\chic{V}}
\widetilde{R}^{\chic{T'}}= -i 2 a_{\chic{V}}  \widetilde{W}^{12}\,.
\eea

For $\widetilde{W}^{\mu \nu}$,
Eqs.~(\ref{hadro},\ref{G},\ref{hadrost}) hold if one replaces $f^{\mu
\nu}$ by
\bea
\la{generPV}
\widetilde{f}^{\mu \nu} =
 &&-\widetilde{w}_1(\tau, \mu^*)\left(g^{\mu \nu}
+\frac{\kappa^\mu \kappa^\nu}{\tau}\right)
   +\widetilde{w}_2(\tau, \mu^*) V^\mu V^\nu \nonumber \\[0.1cm]
 &&-i\widetilde{w}_3(\tau, \mu^*) \epsilon_{\mu \nu \alpha \beta}
\kappa^\alpha V^\beta  \,.
\eea
It is then straightforward to obtain the standard expressions for the
PV response functions, namely
\be
\la{avePV}
\widetilde{R}^{{\chic{L}},{\chic{T}},{\chic{T'}}} (|\vq |, \omega) =  
\int^{\mu^*_{max}}_{\mu^*_{min}} d\mu^* G(\mu^*) 
\widetilde{R}_0^{{\chic{L}},{\chic{T}},{\chic{T'}}}(|\vq |, \omega, \mu^*)    
\ee
and 
\be
\la{R0LTTp}
\widetilde{R}_0^{{\chic{L}},{\chic{T}},{\chic{T'}}}(|\vq |, \omega, \mu^*) = 
\frac{3 {\mathcal N}}{4 |\bkappa| m_{\chic{N}} \eta_{\chic{F}}^3}
\xi_{\chic{F}}\, \Theta (1 - \psi^2) (1 - \psi^2)  
\widetilde{U}^{{\chic{L}},{\chic{T}},{\chic{T'}}}(|\vq |,\omega, \mu^*) \,, 
\ee
where
\bea
\la{ULPV}
\widetilde{U}^{\chic{L}} &=& \frac{\bkappa^2}{\tau} 
\left[ (1+\tau \rho^2) \widetilde{w}_2(\tau,\mu^*)-\widetilde{w}_1(\tau,\mu^*)
+\widetilde{w}_2(\tau,\mu^*){\mathcal D}(\kappa, \mu^*)
\right]\,, \\[0.2cm]
\la{UTPV}
\widetilde{U}^{\chic{T}} &=& 2 \widetilde{w}_1(\tau,\mu^*) 
+\widetilde{w}_2(\tau,\mu^*){\mathcal
D}(\kappa,\mu^*)\,, \\[0.2cm]
\la{UA}
\widetilde{U}^{\chic{T'}} &=& 2 \sqrt{\tau (1+\tau \rho^2)}
\widetilde{w}_3(\tau,\mu^*) \left[ 1 + {\mathcal D}'(\kappa, \mu^*) \right]\,.
\eea
The function ${\mathcal D}(\kappa, \mu^*)$ is defined in Eq.~(\ref{D}), while
\be
\la{D'}
{\mathcal D}'(\kappa, \mu^*)=\frac{1}{|\bkappa |}
\sqrt{\frac{\tau}{1+\tau \rho^2}} \left[1+\xi_{\chic{F}}
(1+\psi^2)+\lambda \rho \right] -1 \,.
\ee
As in the PC case, all of the dynamical information is
carried by the functions $\widetilde{w}_{1,2,3}(\tau,\mu^*)$.

\subsection{Hadronic tensor and form factors}

The PV $N - N^*(1440)$ hadronic tensor arises from the interference
between the electromagnetic and the weak neutral currents
\be
\la{PVhadtensor}
\widetilde{f}^{\mu \nu} = \frac{1}{2} \mu^* \mathrm{Tr}
\left[\frac{(p\!\!\!\!\:/+m_{\chic{N}})}{2 m_{\chic{N}}} \left(\gamma_0 
J_{em}^{\dagger \mu} \gamma_0 \right) \frac{(p\!\!\!\!\:/'+W)}{2 W}
J_{nc}^\nu \right] \,;
\ee
$J_{em}$ is given in Eq.~(\ref{N*curr}) while $J_{nc}$ contains a vector
and an axial-vector part. The vector current has the same structure as 
the electromagnetic one and the axial current is given in terms of 
axial and pseudoscalar form factors just as in the nucleon case:
\be
\la{N*nc}
J_{nc}^\alpha = J_{\chic{V}}^\alpha + J_{\chic{A}}^\alpha 
\ee
with 
\be
\la{JV}
J_{\chic{V}}^\alpha = \bar u_{N^*}(p') \left[ 
\widetilde{F}_1(q^2)\left( q \!\!\!\!\: /\, q^{\alpha} - 
q^2 \gamma^{\alpha}\right) 
+ i \widetilde{F}_2(q^2) \sigma^{\alpha \beta} q_{\beta} \right] u(p)\,,
\ee
\be
\la{JA}
J_{\chic{A}}^\alpha = \bar u_{N^*}(p') \left[ \widetilde{G}_{\chic{A}}
\gamma^\alpha \gamma_5 + \widetilde{G}_{\chic{P}} 
q^\alpha \gamma_5 \right]  u(p)\,. 
\ee
It is convenient to introduce the weak form factors 
$\widetilde{G}_{\chic{E,M}}$ related
to $\widetilde{F}_{1,2}$ exactly as the
electromagnetic form factors in Eqs.~(\ref{GE},\ref{GM}) and it is then
possible to express $\widetilde{G}_{\chic{E,M}}$ in terms of
$G_{\chic{E,M}}$. Following~\cite{Musolf:1994tb} one gets
\be
\la{GEM}
2 \widetilde{G}^{p(n)}_{\chic{E,M}} = (1- 4 \sin^2{\theta_{\chic{W}}})
G^{p(n)}_{\chic{E,M}} - G^{n(p)}_{\chic{E,M}} \,.
\ee
In a similar way, $\widetilde{G}_{\chic{A}}$ can be written as
\be
\la{GA}
2 \widetilde{G}^{p(n)}_{\chic{A}} = G^{p(n)}_{\chic{A}} -
G^{n(p)}_{\chic{A}} = \pm G^{\chic{V}}_{\chic{A}}\,. 
\ee
Lacking any experimental information or model calculations for the $N- N^*(1440)$ 
transition axial-vector form factor $G^{\chic{V}}_{\chic{A}}$,
as in~\cite{Alvarez-Ruso:1998jr} we take
\be
\la{GAV}
G^{\chic{V}}_{\chic{A}}(Q^2) = 2 f_\pi \frac{\tilde{f}}{m_\pi} \left( 1+
\frac{|Q^2|}{M_{\chic{A}}^2} \right)^{-2} \,,
\ee
where the value at $Q^2=0$ has been derived assuming pion pole
dominance in the divergence of the axial current. Here, $f_\pi =
92.4$~MeV is the pion decay constant~\cite{Groom:2000in} and
$\tilde{f}=0.48$ is the $N^* \raw N\,\pi$ coupling (see the Appendix).  
The dipole form for the $Q^2$ dependence with $M_{\chic{A}}\approx 1$~GeV 
is inspired by the analogous behavior of the nucleon axial form factor. 
The pseudoscalar form factor $\widetilde{G}_{\chic{P}}$ can, in
principle, be related to $\widetilde{G}_{\chic{A}}$ if one assumes
partial conservation of the axial current
(PCAC)~\cite{Alvarez-Ruso:1998jr}. However, since for PV electron
scattering we require only the transverse projections of the
axial-vector current, which do not include the pseudoscalar
contributions, the latter are not needed in the present work. For purely
weak interaction processes, and then only when mass terms must be
retained, is it necessary to include such contributions (see, for
example, \cite{Donnelly:1979tz}). 

Finally, substituting the electromagnetic and neutral currents in
Eq.~(\ref{PVhadtensor}) and comparing the result with the general 
form of $\widetilde{f}^{\mu \nu}$ in Eq.~(\ref{generPV}), one obtains
\bea
\la{w1PV}
\widetilde{w}_1 &=& \D   \left[ \tau +\frac{(\mu^* -1)^2}{4} \right]
G_{\chic{M}} \widetilde{G}_{\chic{M}}
\,, \\[0.2cm]
\la{w2PV}
\widetilde{w}_2 &=& \D \frac{1}{\D 1 + \frac{4 \tau}{(\mu^* +1)^2}} 
\left[ G_{\chic{E}} \widetilde{G}_{\chic{E}} + \frac{4 \tau}{(\mu^*
+1)^2} G_{\chic{M}} \widetilde{G}_{\chic{M}}  \right] \,, \\[0.2cm]
\la{w3}
\widetilde{w}_3 &=& \D G_{\chic{M}} \widetilde{G}_{\chic{A}} \,.
\eea

\subsection{Results and discussion}

\begin{figure}[th!]
\begin{center}
\includegraphics[width=\textwidth]{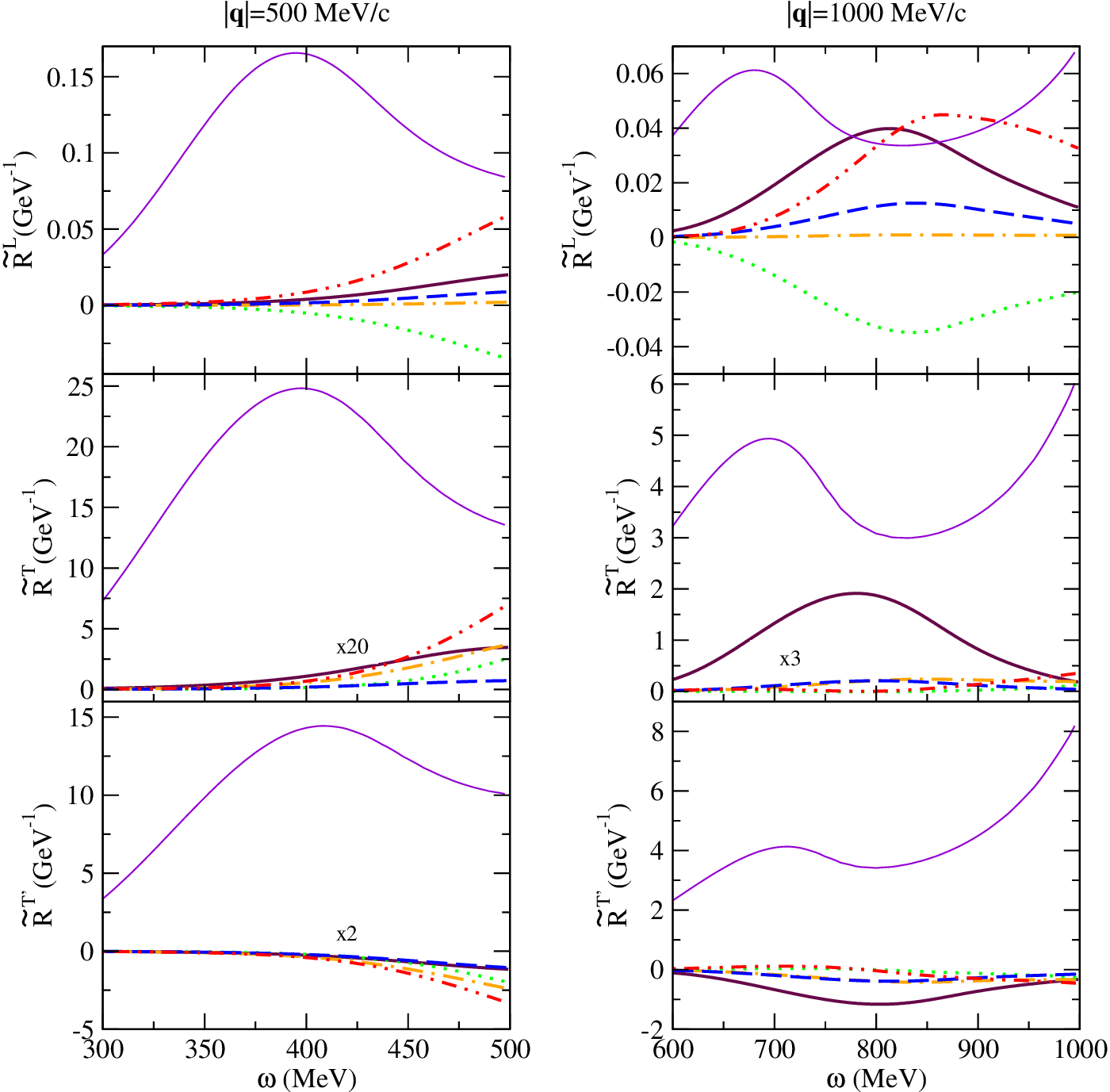}
\caption{PV $N-N^*(1440)$  and $N-\Delta$ (thin solid lines)
responses. In the case of the Roper, the helicity amplitudes of
Fig.~\ref{fig2} have been used; the line styles are the same as in
Figs.~\ref{fig3} and \ref{fig5}. Notice that in some of the plots, the 
$N-N^*(1440)$ responses have been multiplied by factor 20, 3 and 2.}
\label{fig6}
\end{center}
\end{figure}

In Fig.~\ref{fig6} we present the PV $N-N^*(1440)$
response functions together with those for the 
$N-\Delta$~\cite{Amore:2000xm,Alvarez-Ruso:2000bx}, at $|\vq |=500$ and
$1000$~MeV$/c$, and for $^{12}$C ($k_{\chic{F}}=225$~MeV$/c$, $Z=N=6$) as
in the PC case. We have used the helicity amplitudes from the five 
models displayed in Fig.~\ref{fig2}. The Roper contribution to the transverse 
and axial responses is much smaller than in the case of the $\Delta$, so that the
differences between the models do not appreciably change the
results for the full response functions. In the case of
$\widetilde{U}^{\chic{T'}}$, Eq.~(\ref{GAV}) has been used for 
$G^{\chic{V}}_{\chic{A}}$ in all cases. Different model calculations
of these form factors will not change the situation appreciably, since 
the $q^2 =0$ limit is not likely to be very different from the estimate based
on PCAC, while the details of the $q^2$ dependence have little impact
on the response functions, especially at low $|\vq |$.   

The longitudinal response, in spite of its smallness, displays some
interesting features. At $|\vq | = 500$~MeV$/c$, the results obtained
with the NRQM, hybrid and ChD models is small while the LF 
and, particularly,  EVMD
approaches produce sizable (compared with the $\Delta$) responses of
opposite signs. In order to understand the origin of such different
behaviors we have calculated 
the $\tau\raw 0$ limit of $\widetilde{U}^{\chic{L}}$ in terms of the 
helicity amplitudes as in Eq.~(\ref{UL0}), obtaining
\bea
\la{ULPV0}
\widetilde{U}_p^{\chic{L}} (\lambda \raw |\bkappa |) &=&
\left( \frac{m_{\chic{N}}}{2 \pi \alpha} \right) \bigg[ 
8 \bkappa^2 \frac{\mu^{*2}}{\mu^{*2}-1} 
S_{1/2}^p(0) \widetilde{S}_{1/2}^p(0)  \nonumber \\[0.2cm]
&+&   |\bkappa | \xi_{\chic{F}} (1-\psi_0^2)
A_{1/2}^p(0) \widetilde{A}_{1/2}^p(0) \bigg] \,,
\eea
where 
\bea
\la{At}
2 \widetilde{A}_{1/2}^p &=& (1- 4 \sin^2{\theta_{\chic{W}}}) A_{1/2}^p
- A_{1/2}^n \\[0.2cm]
\la{St}
2 \widetilde{S}_{1/2}^p &=&  (1- 4 \sin^2{\theta_{\chic{W}}}) S_{1/2}^p
- S_{1/2}^n \,.
\eea
The contribution arising from the neutrons ($\widetilde{U}_n^{\chic{L}}$) is
obtained by interchanging the labels $p$ and $n$ in the above
equations. As we argued in Sec.~\ref{rad}, the factors in front of the 
helicity amplitudes are such that only the $S_{1/2}
\widetilde{S}_{1/2}$ term matters. Since $(1- 4
\sin^2{\theta_{\chic{W}}})\approx 0.092$, the term proportional to
$S_{1/2}^{p2}(0)$ is disfavored with respect to the one proportional to 
$-S_{1/2}^p(0) S_{1/2}^n(0)$; $S_{1/2}^n(0) \neq 0$ only for LF and
EVMD models (with different signs), as can be seen in
Fig.~(\ref{fig2}). This shows how the nuclear PV response functions mix
the information about the $N-N^*$ transitions on protons and 
neutrons in a non-trivial way.  

\begin{figure}[hd!]
\begin{center}
\includegraphics[width=\textwidth]{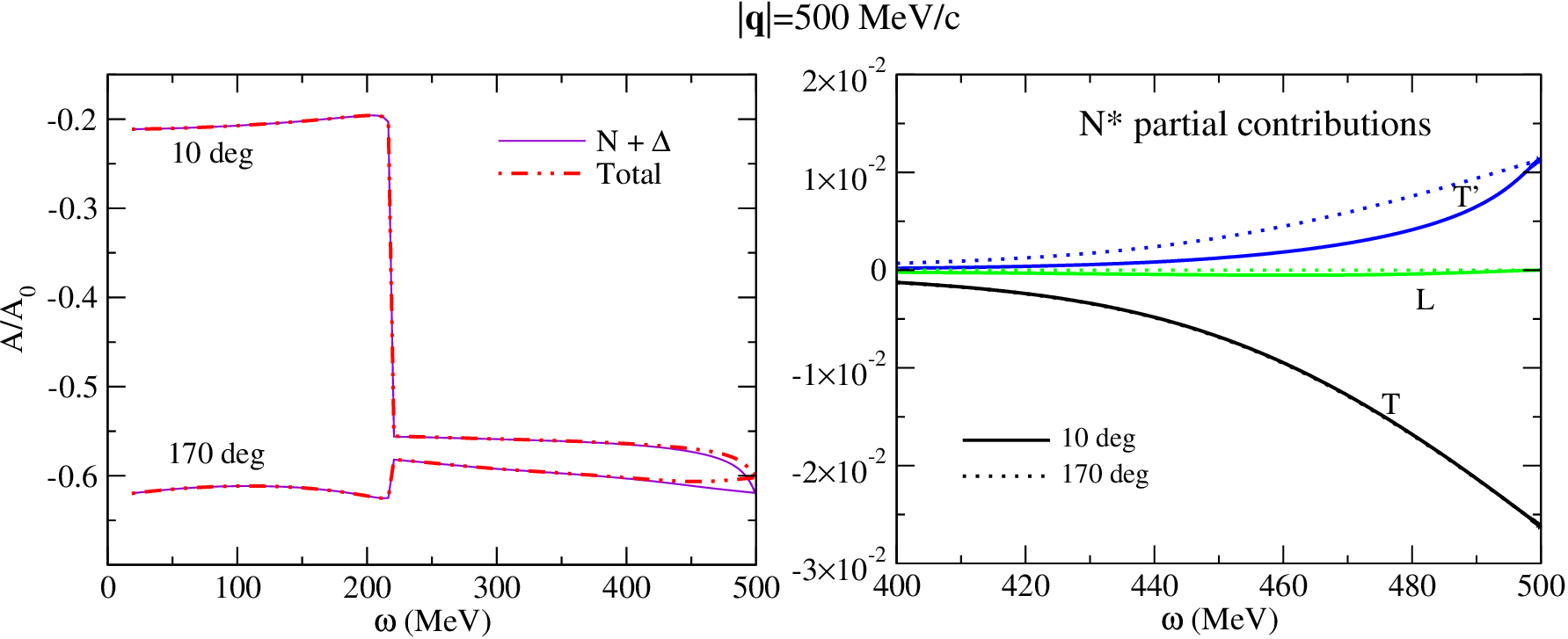} 
\caption{Left panel: the asymmetry in the $N+\Delta$ (solid) 
and  $N+\Delta+N^*$ (dot-dashed) model for $|\vq|=500$ MeV$/c$ at forward and
backward angles; 
right panel: the separate contributions
of the $N^*$ to the asymmetry in the $L$, $T$ and $T'$ channels.}
\label{figasym1}
\end{center}
\end{figure}

\begin{figure}[hd!]
\begin{center}
\includegraphics[width=0.6\textwidth]{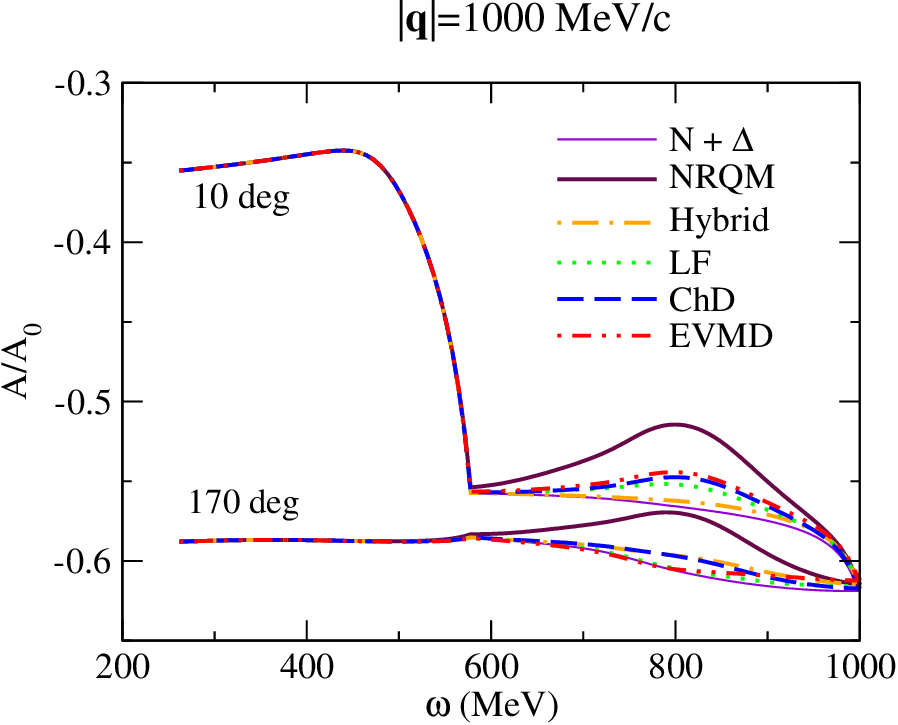} 
\caption{The same as the left panel of Fig.~\protect{\ref{figasym1}},
  but now for 
$|\vq|=1000$ MeV$/c$. The contribution of the Roper is displayed for various
models of the transition amplitudes.}
\label{figasym2}
\end{center}
\end{figure}

In Figs.~\ref{figasym1}, \ref{figasym2} we show
the asymmetry including
the quasielastic contribution, the  $\Delta$ and the $N^*(1440)$
for $|\vq|=500$ and $1000$ MeV$/c$, respectively, and for 
forward (10$^\circ$) and backward
(170$^\circ$) angles. In the case of the lower $|\vq |$ the
contribution of the $N^*(1440)$ to the asymmetry is negligible. Here,
as well as in the case of the PC cross section previously discussed, 
the sizable contribution to $\tilde{R}^L$ 
is completely suppressed by $v_L$ (see the right panel). 
Hence the statement made in~\cite{Amore:2000xm} 
interpreting the deviation from flatness of the
asymmetry as a signal
of the $N - \Delta$ axial response is not affected by the Roper resonance at least
at low momentum transfer. On the other hand, at $|\vq|=1000$~MeV$/c$, the Roper 
contribution
changes the asymmetry above the quasielastic peak appreciably.
Note that in this case the various models considered do not differ much 
from one other; only the NRQM does, and it is in fact one of the least
favored by the data on the proton.
Nevertheless, the striking effect predicted in~\cite{Amore:2000xm}, namely of
an increase in magnitude of the asymmetry of about a factor three 
occurring at forward 
angles in passing from the QEP to the $\Delta$ peak domain, still persists.

\section{Conclusions}

We have studied the nuclear PC and PV response functions for the
$N-N^*(1440)$ transition within the RFG framework. Empirical parameterizations 
and a selection
of the model calculations available in the literature for the electromagnetic
helicity amplitudes have been used as an input. The responses show
some sensitivity to the various models. In particular,
the non-relativistic quark model and the
hybrid model with a large gluon component predict a quite different 
PC longitudinal response $R_{\chic{L}}$. 
Also the $R_{\chic{L}}$ associated with the Roper can be
large compared with the contribution arising from the $\Delta$ near the
light-cone. However, although in principle 
the study of the longitudinal response function in this region could provide
valuable information about the nature of the Roper resonance, in practice
this is hard to achieve because here the longitudinal response is much
suppressed. On the other hand, given a longitudinal/transverse
separation, the impact of 
the Roper on the Coulomb sum rule can be significant, as seen 
in Fig.~\ref{figcsr}.

Our results show that values of the momentum transfer around 500~MeV$/c$ are
the most favorable for studies of $N-N^*(1440)$ transitions because here
the Roper signal is already relatively large due to its considerable width,
while heavier resonances are not likely to be too disruptive. A similar
observation has been recently made in a study of 
photoproduction of the $N^*(1440)$ and vector mesons ($\rho$,
$\omega$)~\cite{Soyeur:2000fm}. 

In spite of the difficulties discussed in the paper with studying this elusive
excitation of the nucleon, the interest in pursuing an in-medium
investigation of the Roper resonance, as we have done here, relates in part to the role
it plays as a breathing mode of the nucleon. It is the analog of the breathing
mode of the nucleus (the isoscalar monopole mode) which carries information on
the compressibility of nuclear matter. The same should
occur at the nucleonic level, and hence a study of the nucleon's 
compressibility, both in free space and embedded in the nuclear medium, may help 
in understanding the nature of hadronic matter including the problem of
confinement.

It should be kept in mind that our treatment of nuclear 
effects is restricted to the inclusion of
Fermi motion; hence it represents only a first step.  More work is
required to take into account polarization effects, meson exchange
currents and  in-medium renormalization in order to treat the
second resonance region with the same level of sophistication achieved for
the $\Delta$ and the quasielastic peaks. 

Finally, it is worth pointing out that the contributions of the 
$N^*(1520)$ and $N^*(1535)$ to the nuclear responses should also be 
accounted for before
drawing too specific conclusions on the role of the Roper. 
Also we have treated the $\Delta$ and
the Roper as independent particles, but since both 
decay mainly into the $N \pi$ channel it is advisable 
to model non-resonant contributions and to estimate the 
importance of interference effects within a consistent framework. The
spirit of the present exploratory study has been to determine whether
or not such an ambitious program is warranted.

\section{Acknowledgments}
We thank J. A. Caballero, C. Maieron, M. Post and M. J. Vicente Vacas for 
useful discussions. This work has been supported in part by the Spanish DGICYT 
contract number BFM2000-1326 and in part (TWD) by funds provided by
the U.S. Department of Energy under cooperative research agreement
No. DE-DC02-94ER40818.

\appendix
\section{The width of the Roper resonance}
\label{ape}

We express the total energy-dependent width of the $N^*(1440)$ as the
sum of three contributions, $N \pi$, $\Delta \pi$ and 
$N (\pi \pi)^{I=0}_{s-wave}$ decay modes,
\be
\la{modes}
\Gamma (W) = \Gamma_{N\pi} (W) + \Gamma_{\Delta \pi} (W) +
\Gamma_{N\pi\pi} (W) \,.
\ee
The more uncertain $N\rho$ channel~\cite{Groom:2000in} has been
neglected. $\Gamma_{N\pi}$ exhibits a p-wave structure which, in the
nonrelativistic limit, can be cast as~\cite{Oset:1985wt} 
\be
\la{Npi}
\Gamma_{N\pi} (W) = \frac{3}{2 \pi} \left( \frac{\tilde{f}}{m_\pi}
\right)^2 \frac{m_{\chic{N}}}{W} |\vq_{cm} |^3 \,,
\ee
where $\vq_{cm}$ is the pion 3-momentum in the resonance rest frame.
The coupling $\tilde{f}=0.48$ is obtained assuming a total width of 350~MeV 
(at $W=m^*$) and an $N\pi$ branching ratio of
65\%~\cite{Groom:2000in}.    

An accurate evaluation of the $N^* \raw \Delta \pi$ width requires one
to
take into account the width of the $\Delta$ resonance. The fact that
the $\Delta$ width is not small compared with the mass difference
between the Roper and the $\Delta$ makes this correction advisable.
The width is then expressed as~\cite{Hirenzaki:1996js}
\be
\la{N*Dpiw}
\Gamma_{\Delta \pi} (W) =  \frac{1}{3 \pi^2} 
\left(\frac{g_{N^* \Delta \pi}}{m_\pi} \right)^2 \int_{0}^{|\vp|_{max}} d|\vp|
\,\frac{\vp^4}{\sqrt{\vp^2 + \mpi^2}} \left| D_\Delta (W_\Delta) \right|^2 
\Gamma_\Delta (W_\Delta)\,,
\ee
where $D_\Delta (W_\Delta)$ is the $\Delta$ propagator
\be
\la{propD}
D_\Delta (W_\Delta)=
\frac{1}{W_\Delta- M_\Delta + \frac{1}{2} i \Gamma_\Delta (W_\Delta)}
\,,
\ee
$\Gamma_\Delta$ is the standard p-wave $\Delta \raw N \pi$ partial
decay width,
\be
W_\Delta^2 = W^2 -2 W (\sqrt{\vp^2 + \mpi^2}) + \mpi^2\,, 
\ee
and
\be
\vp_{max}^2 = \left( 
\frac{W^2 - m_{\chic{N}}^2 - 2 m_{\chic{N}} \mpi}{2 W} \right) ^2 - \mpi^2 \,.
\ee
Using a branching ratio of 25\% one obtains
$g_{N^* \Delta \pi} = 2.07$.

Finally, if we use the simplest possible Lagrangian for the 
$N^* \raw N (\pi \pi)^{I=0}_{s-wave}$ channel as
in~\cite{Oset:1985wt}, the decay width is given by
\be
\Gamma_{N\pi\pi} (W) = 3 c^2 \frac{1}{2^5 \pi^3}
\frac{m_{\chic{N}}}{W^2} \int^{(W-m_{\chic{N}})^2}_{4 m_\pi^2} 
\frac{dx}{x} \lambda^{1/2}(W^2,x,m_{\chic{N}}^2)
\lambda^{1/2}(x,m_\pi^2,m_\pi^2)
\ee
with
\be
\lambda (x,y,z)=x^2 + y^2 +z^2 -2 (x y + x z + y z)\,,
\ee
assuming a branching ratio of 7.5\%, $c=2.3/m_\pi$.  

\bibliography{bibliorespo,biblioauto}

\end{document}